\documentclass[final] {aipproc}
\usepackage{sidecap,graphicx,boxedminipage,epsfig,wrapfig,floatflt,shadow}

\layoutstyle{8x11double}

\begin{document}

\title{Assessing Expertise in Quantum Mechanics using Categorization Task}


\author{Shih-Yin Lin and Chandralekha Singh}{
  address={Department of Physics and Astronomy, University of Pittsburgh, Pittsburgh, PA, 15260, USA}
}

\begin{abstract}
We discuss the categorization of 20 quantum mechanics problems by 6 physics professors and 22 undergraduate students from two honors-level
quantum mechanics courses. Professors and students were asked to categorize the problems based upon similarity of
solution. We also had individual discussions with professors who categorized the problems.
Faculty members' categorizations were overall rated better than those of students by three faculty members who evaluated
all of the categorizations. But the categories created by faculty members were more diverse compared to the uniformity of the 
categories they created when asked to categorize introductory mechanics problems. 

\end{abstract}

\maketitle

\section{Introduction}
\vspace*{-.015in}

A crucial difference between the problem solving strategies used by experts in physics and beginning students lies in the interplay
between how the knowledge is organized and how it is retrieved to solve problems. Categorizing or grouping together problems based upon similarity
of solution can give a glimpse of the ``pattern" an individual sees in a problem while contemplating how to solve it~\cite{chi3,mestre,hardiman}.
In a classic study by Chi et al.~\cite{chi3}, a categorization task was used to assess introductory physics students' level of expertise in physics.
In Chi's study~\cite{chi3}, eight introductory physics students were asked to group introductory mechanics
problems into categories based upon similarity of solution.
They found that, unlike experts (physics graduate students) who categorized them 
based on the physical principles involved to solve them, introductory students 
categorized problems involving inclined planes in one category and pulleys in a separate category~\cite{chi3}.
Previously, we conducted a categorization study in which 21 physics graduate students and several hundred introductory physics students 
were asked to categorize introductory physics problems~\cite{chandra}.
We found that there is a large overlap in the performance of graduate students and introductory
students in the calculus-based courses on the categorization task~\cite{chandra}.

Here, we discuss a study in which 22 physics juniors and seniors (12 and 10 students in the two classes) in two undergraduate quantum 
mechanics courses and six physics faculty 
were asked to categorize 20 quantum mechanics problems based upon similarity of solution. We also discussed issues related to categorization 
individually with faculty members. All but one faculty 
had taught an upper-level undergraduate or graduate level quantum mechanics course. 
The faculty who had not taught quantum mechanics regularly teaches other physics graduate ``core" courses
including E$\&$M and statistical mechanics. 
In each of the two classes, the categorization task was given as a class quiz on the last day of the course and students were given
full quiz credit for their participation.  They were given approximately 35-40 minutes to complete the categorization task.
The faculty members completed the categorization task at a time convenient to them. Except for the faculty member who had not taught
quantum mechanics (who took longer), all the other faculty members noted that it took them less than 30 minutes to complete the categorization task.

The 20 problems to be categorized (given in the Appendix)
were adapted from the problems found at the end of the chapter exercises in commonly used upper-level
undergraduate quantum mechanics textbooks. All those who performed the categorization were provided
the instruction given at the beginning of the Appendix. The sheet
on which individuals were asked to perform the categorization of problems
had three columns. The first column asked them to come up with their
own {}``category name\char`\"{} for each category, the second
column asked them for a description of the category that explains
why those problems can be grouped together and the third column asked
them to list the problem numbers for the problems that should be placed in that category.

The goal was to investigate the differences in categorization
by faculty members and students and whether there are major differences in the ways in which individuals in each group categorize quantum mechanics 
problems. This study was partly inspired by the fact that a physics faculty 
who was teaching advanced 
undergraduate quantum mechanics in a previous semester had given a take-home exam in which one problem asked students to find the wave function of a 
free particle after a time $t$ given the initial wave function (which was a Gaussian). Two students approached the faculty member complaining that this 
material was not covered in the class. The faculty member pointed out to them that he had discussed in the class how to find the wave function 
after a time $t$ given an initial wave function in the context of a problem involving an infinite square well. But the students insisted that, while the 
time-development of the wave function may have been discussed in the context of an infinite square well, it was not discussed in the context of a free particle. It 
appears that the two students did not categorize the time-development issues for the infinite square well and the free particle in the same category. 
They did not realize that a solution procedure very similar to what they had learned in the context of the time-development of the wave function for an 
infinite square well should be applicable to the free particle case except they must use the energy eigenstates and eigenvalues corresponding to the 
free particle and replace the discrete sum over energy levels for an infinite square well by an integral since the energy levels for a free particle 
are continuous.

\vspace*{-.231in}
\section{Scoring of Categorization}
\vspace*{-.11in}

We find that the categorizations of a problem performed by the students were diverse. But the faculty members also categorized each problem in
categories that were often different, unlike the uniform categorization by faculty members for introductory physics problems~\cite{chandra}.
To analyze the data quantitatively, we placed each category created by each individual into a matrix which consisted of 
problem numbers along the columns and categories along the rows. A ``1" was assigned if the problem appeared in the given 
category and a ``0" was assigned if the opposite was true. 
Categories that were very similar were combined, e.g., ``time-dependence of wave function" and "time-development of wave function" or ``dynamics of wave function" were combined into a single category.
In order to score the categorizations by students and faculty members, three faculty members (a subset of those who had categorized the problems 
themselves) were recruited. They were given the categorizations by students and faculty in the matrix form we had created 
(without identifiers and with the categorizations by the faculty and students jumbled up). All the different categories created by different 
individuals for problem (1) were placed one after the other to aid faculty members who were scoring the categorizations. For each question,
faculty members doing the scoring were advised to read the question, think about how they would categorize it and then evaluate and score
everybody's categorization.
The three faculty members were asked to evaluate whether 
each of the categories created by an individual should be considered {}``good\char`\"{} (assigned a score of 2), {}``moderate\char`\"{} (assigned
a score of 1), or {}``poor\char`\"{} (assigned a score of zero). We note that if all three faculty members scored a particular problem for an 
individual as ``good", the score of that individual on that problem will be 6 (maximum possible). If one faculty 
scored it as ``good" but the other two scored it as ``medium", the score of that individual on that problem will be 4. 

\vspace*{-.235in}
\section{Results}
\vspace*{-.11in}

Each of the 22 students and 6 faculty members categorized 20 problems. 
Some of the categories for a problem were common for more than one individual and sometimes an individual placed a problem in more than one category. 
All the three faculty members
noted that evaluating and scoring other people's categorization was a very challenging task and required intense focus. One faculty member noted
that it took him several hours to complete the scoring. Moreover, two of the faculty members who evaluated everybody's categorization noted that they 
would prefer not to use the terms ``good" or ``poor" for judging the categories.
The faculty members who scored others' categorizations also noted that 
sometimes they liked the categorizations of a problem by others much more than their own.
Interestingly, in our earlier studies with introductory physics categorization, we had asked three faculty members to evaluate the
categorizations of a subset of randomly selected individuals (in that case we did not ask them to score all categorizations because the introductory
physics classes had several hundred students)~\cite{chandra}. In scoring introductory physics categorizations, faculty 
were not hesitant about calling the categories good/poor and they did not say that the task was challenging~\cite{chandra}. 
They also never said that they preferred others' categorizations
of a problem more than their own perhaps because there was a great conformity in faculty categorizations (which were based upon physics principles such as 
the conservation of mechanical energy, conservation of momentum, conservation of angular momentum, Newton's second law etc.)~\cite{chandra}.

\begin{figure}[h!]
\epsfig{file=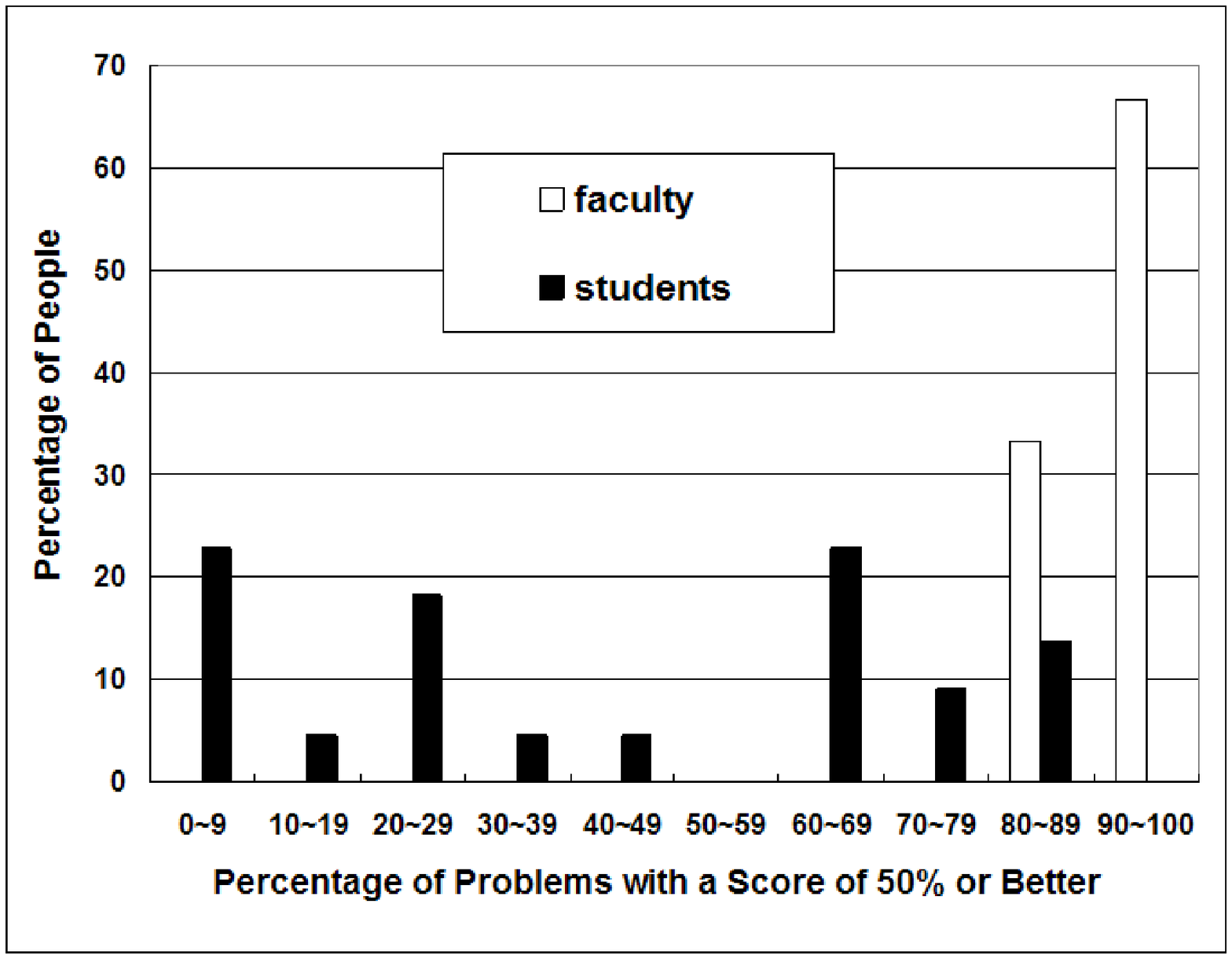,height=2.28in}
\epsfig{file=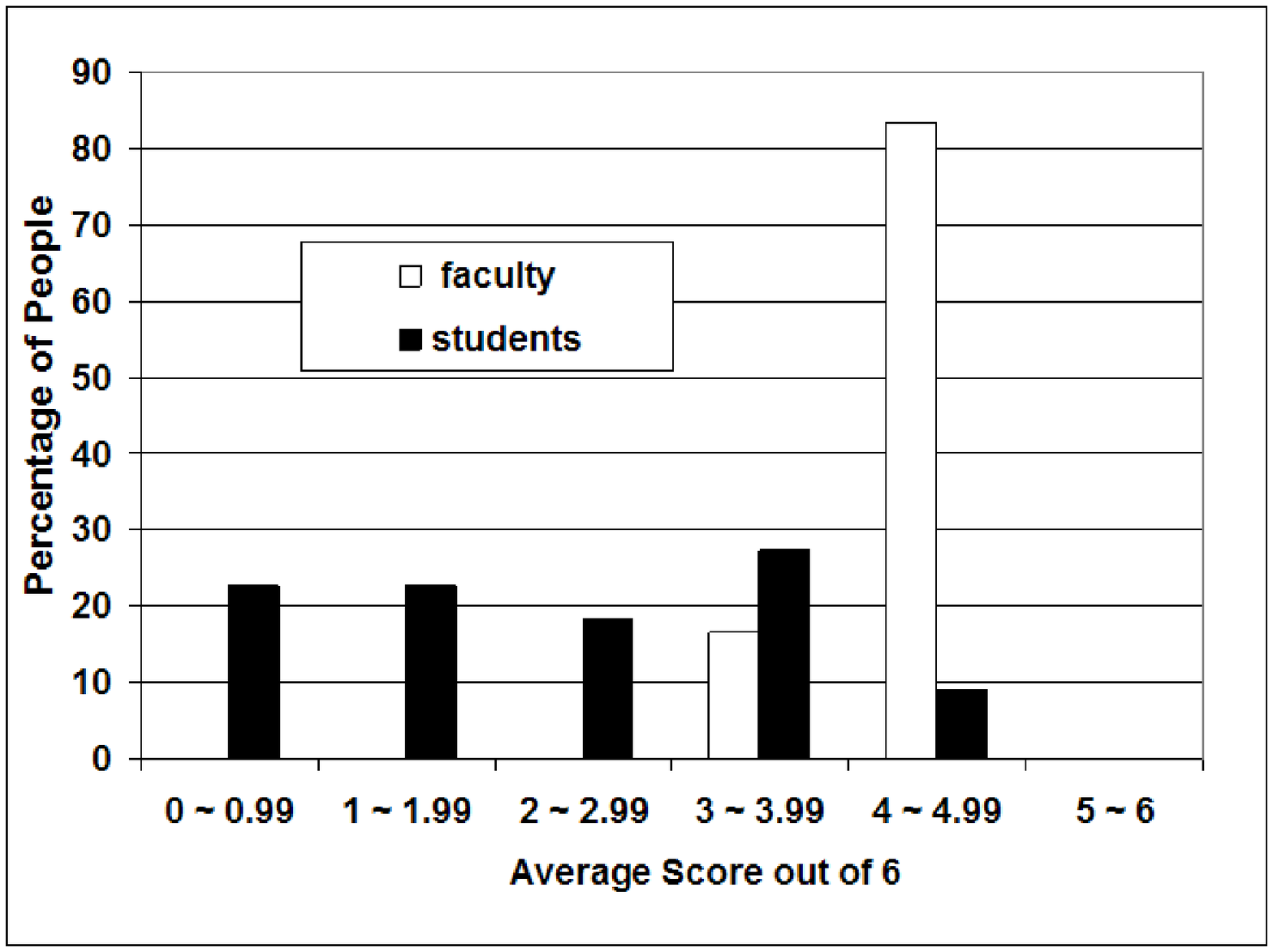,height=2.28in}
\caption{
(a) Percentage of people vs. percentage of problems with a score of $50\%$ of better (at least 3 out of 6), (b) 
Percentage of people vs. Average score out of 6}
\end{figure}

Figure 1 (a) shows a histogram of the percentage of people (students or faculty) vs. percentage of problems with a score of $50\%$ of better 
(at least 3 out of 6) and Figure 1 (b) shows a histogram of the percentage of people vs.
average score on the categorization task out of a maximum of 6 (averaged over all problems). 
We note that what one faculty member scored as ``good" was often scored as ``medium" by another.
As examples, for problem (2), ``time-dependence of expectation value" or ``evolution of states" received a score of 6 each, ``angular momentum",
``eigenvalue, eigenfunctions" or
``stationary state" received a score of 4 out of 6 each. For question (13), ``effects of measurement",
``energy and position measurement" or ``collapsed wave function" received a score of 5 each, and ``scalar product", 
``wavefunctions and probability" or ``eigenvalue" received a score of 3 each. For question (14), ``time evolution of wave function" received a score 
of 6, ``superposition"
received a score of 5, ``stationary state" or ``time dependent Schroedinger equation" received a score of 4 each, 
``hydrogen atom" or ``time dependent vs. time independent" received a score of 3 each. For question (15), ``time dependent expectation value",
``expectation value of observable", ``selection rules", ``stationary state", ``symmetry (even/odd)" or ``time dependency, evolution" 
received a score of 4 each and ``time evolution of wavefunction" or ``probability and expectation value"
received a score of 3 each. For question (20), ``effects of measurement",
or ``collapsed wavefunction" received a score of 6, ``hydrogen atom" or ``matrix element" 
received a score of 3.

Figure 1 shows that the categorizations by faculty 
were rated higher overall than those by students, despite the diversity
in faculty responses. Individual discussions with faculty 
suggest that some felt that the structure of knowledge in quantum mechanics 
is more complex than that in introductory physics
which may make it difficult to focus on the same aspects of solution when asked to categorize.
One possible implication is that the way concepts are emphasized in a quantum mechanics course
may differ based upon the ``patterns" that appear to be most central to the faculty member teaching the course.
Another common theme that emerged is that categorization
of introductory physics problems involves identifying fundamental principles relevant for problems, whereas in undergraduate
quantum mechanics it mainly involves
identifying concepts and procedures, because problem solving is tied to conceptual and procedural knowledge.

The overall scores (by the three faculty members who evaluated all of the categorizations) of concrete categories such as
``hydrogen atom" or ``harmonic oscillator" were higher than other concrete categories such as ``infinite square well" or ``free particle", where
four questions out of 20 given in the categorization task belong to each of these four systems. 
Discussions with individual faculty suggest that
they have a notion of a canonical quantum system that they use for thinking about concepts and to help clarify ideas about quantum mechanics.
``Hydrogen atom" and ``harmonic oscillator" fit their notion of canonical quantum systems. One faculty member explicitly noted that the hydrogen
atom and harmonic oscillator are quintessential in quantum mechanics. He added that the hydrogen atom embodies many essential features of other complex
quantum systems but is simple enough to be analysed easily. Similarly, the harmonic oscillator is used as a model to understand diverse quantum systems.
Such explanations about why the average score for ``hydrogen atom" as a category was at least $50\%$ (3 out of 6 including the scores
of all the three faculty who evaluated the responses) for three of 
the four questions that related to the hydrogen atom but was not $50\%$ for any of the four questions related to the infinite square well
shed some light on why the faculty members do not view all ``concrete" categories on the same footing. 

\vspace*{-.24in}
\section{Summary}
\vspace*{-.13in}

The categorization task is useful in that it reflects the salient points and patterns an individual sees in a problem when 
contemplating how to solve it. 
Even in the context of categorization of quantum mechanics problems, professors overall performed better than students as evaluated by
three faculty members who scored everything without the knowledge of which ones belonged to the faculty or students.
Another finding is that, unlike the categorization of introductory problems, in which
professors' categorizations are 
uniform, their categorizations were more varied in the context of quantum
mechanics.

\vspace*{-.24in}
\begin{theacknowledgments}
We thank NSF for awards PHY-0653129 and 055434.
\end{theacknowledgments}
\vspace*{-.14in}

\bibliographystyle{aipproc}

\begin{center}
{\bf Appendix: Categorization Questions} 
\end{center}
$\bullet$ {Your task is to group the 20 problems below into various groups based upon similarity of solution on the sheet of paper provided. 
You can create as many categories as you wish. The grouping of problems should NOT be in terms of
``easy problems", ``medium difficulty problems" and ``difficult problems" but rather it should be based upon the features
and characteristics of the problems that make them similar.  A problem can be placed in more than one group created by you.
Please provide a brief explanation for why you placed a set of questions in a particular group. You need NOT solve any problems.}\\

\noindent
{The first TWO questions refer to the following system:
An electron is in an external magnetic field B which is pointing
in the z direction. The Hamiltonian for the electron spin is given by $\hat H=-\gamma B \hat S_z$ where $\gamma$ is the
gyromagnetic ratio and $\hat S_z$ is the z component of the spin angular momentum operator.\\
}

\begin{enumerate}
\item
If the electron is initially in an eigenstate of $\hat S_x$,
does the expectation value of $\hat S_x$ depend on time?  Justify your answer.

\item
If the electron is initially in an eigenstate of $\hat S_z$, does the expectation value of $\hat S_x$ depend on time?
Justify your answer.

\item
A free particle has the initial wave function
$\Psi(x, t = 0) = Ae^{-ax^2}e^{i k_0 x}$
where $A$, $a$, and $k_0$ are constants ($a$ and $k_0$ are real and positive).
Find $|\Psi(x,t)|^2$.

\item
A particle in an infinite square well ($0\le x \le a$) has the initial wave function $\psi(x,0)=A x (a-x)$.
Find the uncertainty in position and momentum.

\item
In the ground state of the harmonic oscillator, what are the expectation values of position, momentum and energy?
Do these expectation values depend on time?

\item
A particle is in the first excited state of a harmonic oscillator potential. Without any calculations, explain
what the expectation value of momentum is and whether it should depend on time.

\item
A free particle has the initial wave function
$\Psi(x, t = 0) = A e^{ik_0 x}$
where $A$, and $k_0$ are constants ($k_0$ is real and positive).
Find $|\Psi(x,t)|^2$.

\item
An electron is in the ground state of a hydrogen atom.
Find the uncertainty in the energy and the $z$ component of angular momentum.

\item
Make a qualitative sketch of a Dirac delta function $\delta (x)$. Then, make a qualitative sketch of the absolute
value of the Fourier transform
of $\delta (x)$. Label the axes appropriately for each plot.

\item
A free particle has the initial wave function
$\Psi(x, t = 0) = Ae^{-ax^2}e^{i k_0 x}$
where $A$, $a$, and $k_0$ are constants ($a$ and $k_0$ are real and positive).
Find $\langle x\rangle$, $\langle p\rangle$, $\langle x^2\rangle$, $\langle p^2\rangle$,
$\sigma_x=\sqrt{\langle x^2\rangle- \langle x\rangle^2}$, $\sigma_p=\sqrt{\langle p^2\rangle-\langle p\rangle^2}$.

\item
An electron in a hydrogen atom is in a linear superposition of the first and third excited states. Does the expectation
value of its kinetic energy depend on time?

\item
Suppose that the measurement of the position of a particle in an infinite square well ($0\le x \le a$) yields the value $x=a/2$ at the center of the well.
Show that if energy is measured immediately after the position measurement, it is equally probable to find the particle in any odd-energy
stationary state.

\item
An electron is in a linear combination of the ground and fourth excited states in a harmonic oscillator potential.
A measurement of energy is performed and then followed by a measurement of position. What can you say about the possible
results for the energy and position measurements?

\item
An electron in a hydrogen atom is in a linear superposition of the first and third excited states.
Find the wave function after time $t$.

\item
A particle is in the third excited state of a harmonic oscillator potential. Without any calculations, explain what the
expectation value of momentum is and whether it should depend on time.

\item
A particle in an infinite square well ($0\le x \le a$) has the initial wave function $\psi(x,0)=A x (a-x)$.
Without normalizing the wave function, find $\psi(x,t)$.

\item
A free particle has the initial wave function
$\Psi(x, t = 0) = A e^{ik_0 x}$
where $A$, and $k_0$ are constants ($k_0$ is real and positive).
Find $\langle x\rangle$, $\langle p\rangle$, $\langle x^2\rangle$, $\langle p^2\rangle$,
$\sigma_x=\sqrt{\langle x^2\rangle- \langle x\rangle^2}$, $\sigma_p=\sqrt{\langle p^2\rangle-\langle p\rangle^2}$.

\item
A particle is initially in a linear combination of the ground state and the first excited state of an infinite square well.
Without any calculations, explain whether the expectation value of position should depend on time.

\item
What is the commutation relation [$\hat S_x,\hat S_y$]?

\item
A hydrogen atom is in the first excited state. You measure the distance of the electron from the nucleus first and then
measure energy. Describe the possible values of energy you may measure.

\end{enumerate}

\end{document}